\documentclass[aip,reprint,showpacs,twocolumn,superscriptaddress,floatfix]{revtex4-1}
\usepackage{epsfig}
\usepackage{amsmath}
\usepackage{amssymb}
\usepackage{amsfonts}
\usepackage{mathptmx}
\usepackage{dcolumn}
\usepackage{eucal}
\usepackage{bm}
\usepackage{color}
\usepackage[colorlinks,linkcolor=blue,citecolor=blue]{hyperref}

\usepackage{epstopdf}

\usepackage{graphicx}

\usepackage{natbib}

\begin{document}
\title{ Thermal rectification of electrons in hybrid normal metal-superconductor nanojunctions}
\author{F. Giazotto}
\email{giazotto@sns.it}
\affiliation{NEST, Instituto Nanoscienze-CNR and Scuola Normale Superiore, I-56127 Pisa, Italy}
\author{F. S. Bergeret}
\email{sebastian\_bergeret@ehu.es}
\affiliation{Centro de F\'{i}sica de Materiales (CFM-MPC), Centro
Mixto CSIC-UPV/EHU, Manuel de Lardizabal 4, E-20018 San
Sebasti\'{a}n, Spain}
\affiliation{Donostia International Physics Center (DIPC), Manuel
de Lardizabal 5, E-20018 San Sebasti\'{a}n, Spain}

\begin{abstract}
We theoretically investigate heat transport in hybrid normal metal-superconductor (NS) nanojunctions focusing on the effect of thermal rectification. We show that the heat diode effect in the junction strongly depends on the transmissivity and the  nature of the NS contact. 
Thermal rectification efficiency can reach up to $\sim 123 \%$ for a fully-transmissive \emph{ballistic} junction and up to $84\%$ in \emph{diffusive} NS contacts. Both values   exceed the rectification efficiency  of a NIS \emph{tunnel} junction (I stands for an insulator) by a factor close  to $\sim 5$ and $\sim 3$, respectively.
 Furthermore, we show that for NS point-contacts  with low transmissivity,  \emph{inversion} of the heat diode effect can take place. 
Our results could prove  useful for tailoring heat management at the nanoscale, and for mastering thermal flux propagation in low-temperature caloritronic nanocircuitry.    
\end{abstract}

\maketitle

Control of the heat  flow at the nanoscale has been attracting the attention of several research groups in the last decade.\cite{Giazotto2006, Dubi2011}
An accurate understanding of  heat transport is essential, for instance, for a fine  control of  ultrasensitive cryogenic radiation detectors \cite{Giazotto2006,Giazotto2008},  nanocoolers \cite{Giazotto2006,Muhonen2012} and  caloritronic circuits.\cite{Maki1965,Guttman97,Zhao2003,Meschke2006,Chandrasekhar2009,Ryazanov1982,virtanen2007}. In several cases such devices contain superconductors as building block elements 
which introduce phase coherence to the heat transport. 
Examples include Josephson heat interferometers \cite{giazottoexp2012}  and thermal quantum diffractors \cite{Martinezexp2013} in which the heat current is controlled by  a magnetic flux, or
 electronic refrigeration in normal metal-superconductor (NS) \cite{Giazotto2006} and ferromagnet-superconductor (FS) \cite{Giazotto2002,Ozaeta2012} structures 
 whose efficiency depends on  Andreev reflection \cite{Andreev1964} at the  interface with the superconductor. 
  
 In a voltage-biased NS junction the charge current consists of two contributions: the quasiparticle and the Andreev current \cite{Andreev1964}. 
For  voltages $V$ below the superconducting energy gap the latter may dominate, and the amplitude of the current depends on the transmisivity and the nature of the contact. 
Due to the electron-hole symmetry and for a spatially-symmetric barrier at the SN interface,  the amplitude of the electric current does not depend on the sign of $V$.  The same holds as well for the heat current flowing through the junction in a voltage-biased configuration \cite{Giazotto2006}.   
By contrast, the electronic contribution to the heat current in the presence of a temperature bias across the NS junction depends on the sign of the temperature drop \cite{MartinezRect2013}. This property stems from the strong temperature dependence of the superconducting density of states at high temperatures. In this regard, a NS junction therefore behaves as a \emph{thermal diode} \cite{Roberts2011,Casati2007} with this meaning that heat conduction along one direction is preferred with respect to that occurring upon temperature bias reversal.
Strong effort has been devoted so far to envision and to realize thermal rectifiers dealing, for instance, either with phonons \cite{Wu2009,Segal2008,Li2006,Terraneo2002,Chang2006}, electrons\cite{MartinezRect2013,Ren2013,Ruokola2011,Kuo2010,Ruokola2009,Chen2008} or with photons.\cite{age}
\begin{figure}[h]
\includegraphics[width=\columnwidth]{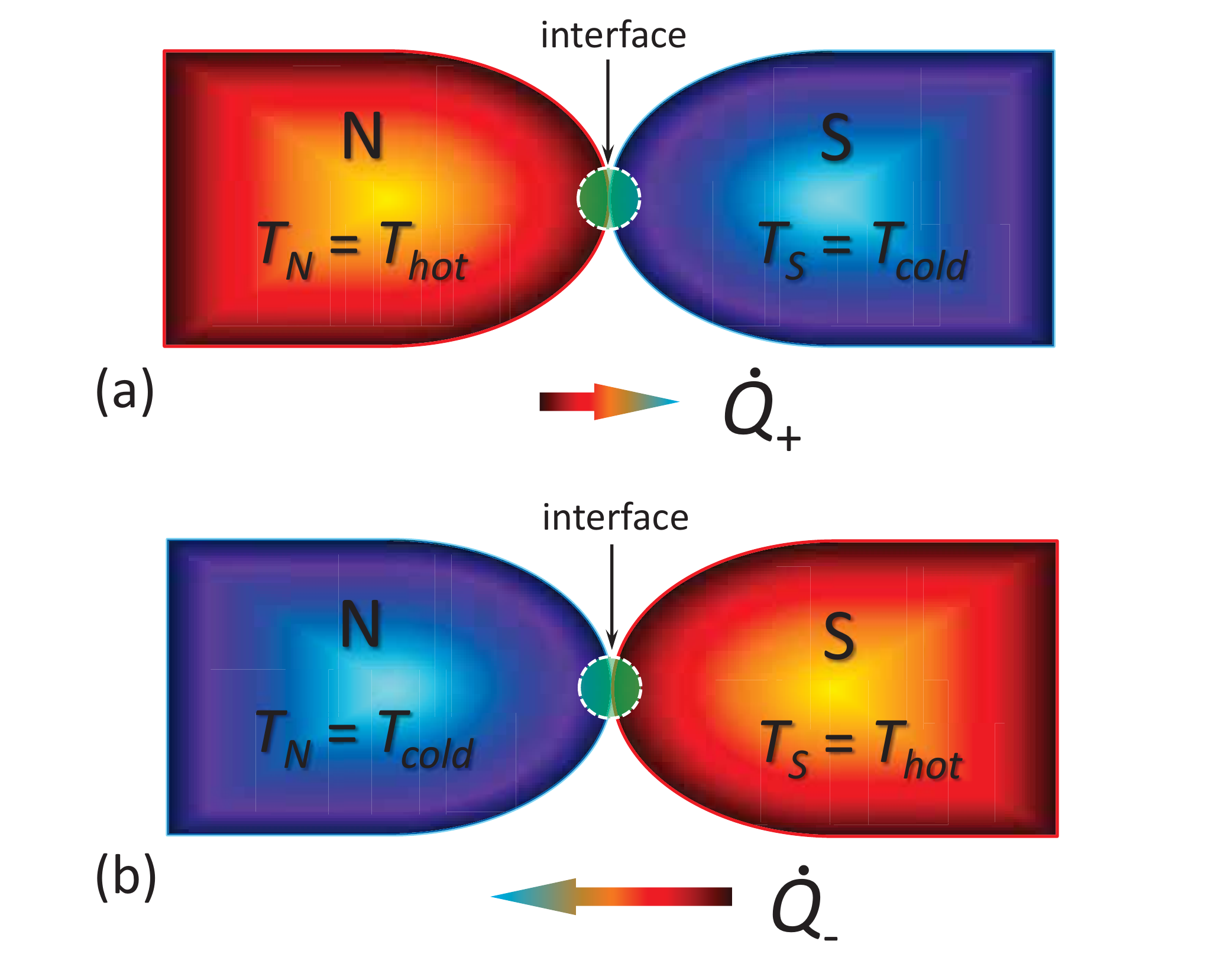}
\vspace{-6mm}
\caption{Scheme of a hybrid normal metal-superconductor (NS) heat diode under forward, (a), and reverse, (b), thermal bias configuration. The NS junction is temperature-biased with $T_N\neq T_S$, and $\dot{Q}_+$ and $\dot{Q}_-$ denote the heat current flowing through the structure in the forward ($T_N>T_S$) and reverse ($T_N<T_S$) thermal-bias setup, respectively. The circular hatched regions indicates the NS interface which, as discussed in the text, can describe a \emph{ballistic} or \emph{tunnel} junction as well as a \emph{diffusive} or \emph{dirty} contact.}
\label{fig1}
\end{figure}

In this Letter we  address the heat diode effect in NS nanojunctions and explore how thermal rectification depends on the  interface properties. 
 We show that a perfectly-transparent point contact can provide a large rectification coefficient up to $\sim 123\%$ which exceeds by a factor close to $5$  the one 
 predicted to occur in NIS tunnel junctions \cite{MartinezRect2013}. In more realistic   diffusive junctions  the maximum heat rectification efficiency can be as large as $\sim 84\%$.
 Furthermore, in a NS point-contact  thermal rectification can change sign as a function of temperature for a low enough interface transmissivity.
  Our predictions for the heat diode effect in hybrid NS junctions could prove useful for  
developing future caloritronic nanodevices.

The system under investigation is schematized in Fig. \ref{fig1} and consists of a temperature-biased  NS junction.
The electronic temperature in N and S is set to $T_N$ and $T_S$, respectively. 
We assume a  spatially uniform temperature in the  electrodes so  to avoid the generation of any thermal gradient within each of them.
In the \emph{forward} thermal bias configuration [see Fig. \ref{fig1}(a)] a thermal gradient is intentionally created developing at the NS interface by setting $T_N=T_{hot}>T_S=T_{cold}$ which gives rise to a total heat flux $\dot{Q}_+$ through the system. 
By contrast, in the \emph{reverse} thermal bias configuration the thermal gradient is inverted so that $T_N=T_{cold}<T_S=T_{hot}$   which yields a total heat current $\dot{Q}_-$ flowing from S to N [see Fig. \ref{fig1}(b)]. 
We note that by definition $\dot{Q}_+$ and $\dot{Q}_-$ have opposite sign. 
The  hatched circles in the figure indicate the NS contact region which, as discussed below, can be ballistis or tunnel as well as diffusive or dirty. The thermal rectification coefficient ($R$) can be defined as\cite{MartinezRect2013}
\begin{equation}
R(\%)=\frac{|\dot{Q}_-|-\dot{Q}_+}{\dot{Q}_+}\times 100.
\label{rect}
\end{equation}
According to Eq. (\ref{rect}),  $R=0$ implies the absence of a heat rectification 
whereas $R>0$ implies a thermal current flowing preferentially from the S toward the N side of the junction. 
  
For a quantitative description of the charge  and heat transport through the NS junction it is convenient to introduce the Keldysh Green's functions
\begin{equation}
\check G_{S(N)}=\left( \begin{array}{cc}
\hat G^R&\hat G^K\\
0&G^A
\end{array}
\right),
\end{equation}
where the  retarded (R), advanced (A) and Keldysh  components in the S and N electrodes are given by 
\begin{eqnarray}
\hat G^{R(A)}_N&=&\pm\tau_3\label{GF}\\
\hat G^K_N&=&2\tau_3\tanh\left(\frac{E}{2k_BT_N}\right)\\
\hat G^{R(A)}_S&=&g^{R(A)}\tau_3+f^{R(A)} i\tau_2\\
\hat G^K_S&=&(\hat G^R_S-\hat G^A_S)\tanh\left(\frac{E}{2k_BT_S}\right).\\
\end{eqnarray}
In the above expressions, $\tau_i$ are the Pauli matrices in the Nambu space, $g^{R(A)}=(E/\Delta(T_S))f^{R(A)}=E/\xi^{R(A)}$, $\xi^{R(A)}=\sqrt{(E\pm i\eta)^2-\Delta^2(T_S)}$, $\Delta(T_S)$ is the BCS temperature-dependent superconducting order parameter, $T_{N(S)}$ is the temperature of the N (S) electrode, and $k_B$ is the Boltzmann constant. The parameter $\eta$ accounts for the inelastic scattering rate within the relaxation time approximation \cite{Dynes1984,Pekola2004}.

\begin{figure}[tb]
\includegraphics[width=\columnwidth]{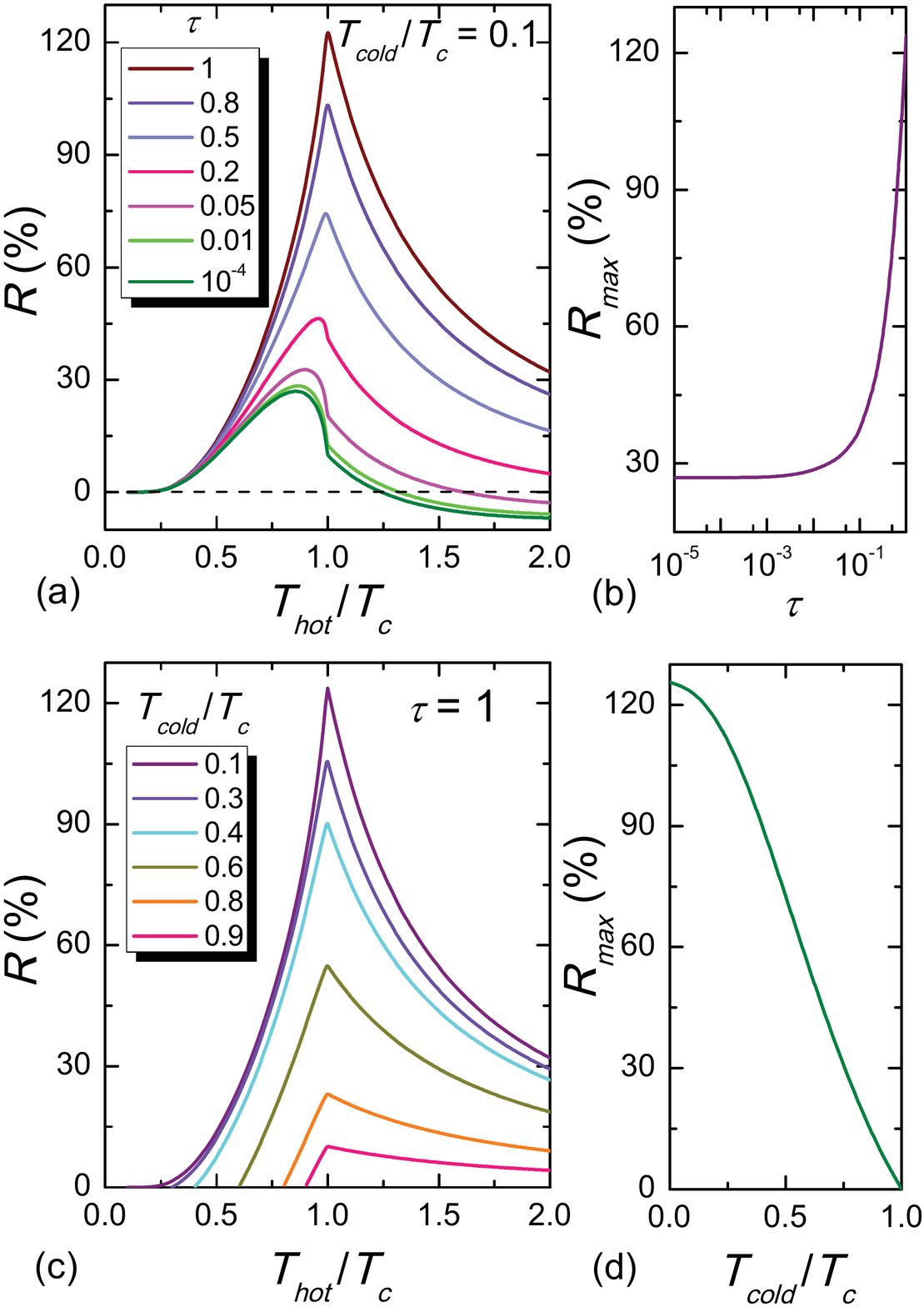} 
\caption{(a)  Rectification efficiency of a point-contact NS junction  $R$ vs $T_{hot}$ calculated at $T_{cold}=0.1 T_c$ for several values of the transmission coefficient $\tau$.
(b) Maximum thermal rectification efficiency of a ballistic junction $R_{max}$ vs $\tau$ for $T_{cold}=0.1 T_c$.
(c) $R$ vs $T_{hot}$ for a perfectly-transmitting ($\tau=1$) ballistic NS junction calculated for a few values of $T_{cold}$.
(d) $R_{max}$ vs $T_{cold}$ calculated for $\tau=1$.
$T_c$ denotes the superconducting critical temperature.
}
\label{fig2}
\end{figure}
The electronic transport through the NS junction can be described using the matrix current ($\hat I$) introduced by Nazarov, \cite{Nazarov99}
\begin{equation}
\hat I=-\frac{2e^2}{\pi\hbar}\sum_n\tau_n\left[\check G_S,\check G_N\right]\left[4-\tau_n\left(2-\left\{\check G_N,\check G_S\right\}\right)\right]^{-1}\; .\label{I}
\end{equation}
 Here,  $\tau_n$ is the transmission of the $n$th junction channel, and the sum goes over the junction conducting channels.  In our analysis we shall focus on the electronic contribution to the heat current only, $\dot Q$, which is defined as
\begin{equation}
\dot Q=\frac{1}{8e^2}\int_{-\infty}^{\infty} E{\rm Tr}\hat  I^K {\rm d}E,
\label{Qdot}
\end{equation} 
and we do not take into account neither  the heat exchanged between electrons and phonons nor a pure phononic heat current \cite{Maki1965,giazottoexp2012}.
From Eqs. (\ref{GF}-\ref{Qdot})  we get for $T_N\neq T_S$ the following expression for the heat current flowing through the contact
\begin{eqnarray}
\dot Q&=&\frac{1}{2\pi\hbar }\sum_n \int_{-\infty}^{\infty} {\rm d} EE\frac{\tau_n}{2-\tau_n(1+g^A)}  \nonumber\\
&\times& \left[(g^R-g^A)-\frac{2\tau_n(f^R-f^A)f^R}{4-2\tau_n(1-g^R)}\right] \nonumber\\
&\times& \left[\tanh\left(\frac{E}{2k_BT_S}\right)-\tanh\left(\frac{E}{2k_BT_N}\right)\right].
\label{finQdot}
\end{eqnarray}
In our notation $\dot Q>0$ represents the heat current flowing out of the  N  lead when $T_N>T_S$.   
Equation (\ref{finQdot}) is a general expression that describes the heat flow for an arbitrary contact. 
For example, a point-contact is defined by a unique conducting channel with transmission $\tau$.  
A \emph{ballistic} junction is described  by setting all channel transmissions $\tau_n=1$, whereas in the case 
 of a   \emph{tunnel} contact all  $\tau_n\ll 1$. 
In the latter case, from Eq. (9) we recover the well-known expression for the heat current ($\dot{Q}_{tunnel}$) flowing through a temperature-biased superconducting \emph{tunnel} junction \cite{Giazotto2006}, i.e., 
\begin{eqnarray}
\dot {Q}_{tunnel}&=&\frac{G_N}{e^2}\int_{\Delta(T_S)}^{\infty} {\rm d} E\frac{E^2}{\sqrt{E^2-\Delta(T_S)^2}}\label{tunnel}
\\
&\times&\left[\tanh\left(\frac{E}{2k_BT_S}\right)-\tanh\left(\frac{E}{2k_BT_N}\right)\right]\nonumber,
\end{eqnarray} 
where $G_N=(e^2/\pi\hbar)\sum_n\tau_n$ is the contact normal-state conductance.

In the case of an extended NS interface with a continuous distribution of channels one can replace in Eq. (\ref{finQdot}) the sum $\sum_n$ with the integral $\int_0^1 d\tau \mathcal P(\tau)$, where $\mathcal P(\tau)$ is the interface transmission distribution function. Realistic interfaces between metals are typically dirty, and can be described by a scattering region of a certain characteristic length. If this length is larger than the Fermi wave length, the interface is called to be \emph{diffusive},  and  is characterized by the following 	distribution function 
\cite{Dorokhov82}
\begin{equation}
\mathcal P(\tau)=\frac{\hbar G_N}{2e^2}\frac{1}{\tau \sqrt{1-\tau}}. 
\label{diffusive}
\end{equation}
By contrast, if the characterisc scattering region is much smaller than the Fermi wave length (i.e., a sharp interface), the distribution function reads\cite{Shep97}
\begin{equation}
\mathcal P(\tau)=\frac{\hbar G_N}{e^2}\frac{1}{\tau^{3/2} \sqrt{1-\tau}}. 
\label{dirty}
\end{equation}
Thus, with the help of Eqs. (\ref{finQdot},\ref{diffusive},\ref{dirty}) we are able to describe heat transport through a large  variety of junctions and obtain the thermal rectification coefficient $R$. 
In the normal state, {\it i.e.},  for temperatures larger than the superconducting critical one, 
 $\Delta =0$ and Eq. (\ref{finQdot}) reduces to 
\begin{equation}
\dot Q=\frac{k_B^2G_N\pi^2}{6 e^2}(T_N^2-T_S^2).
\end{equation}
This expression  shows that no thermal rectification  occurs in a full normal-metal junction.
\begin{figure}[tb]
\includegraphics[width=\columnwidth]{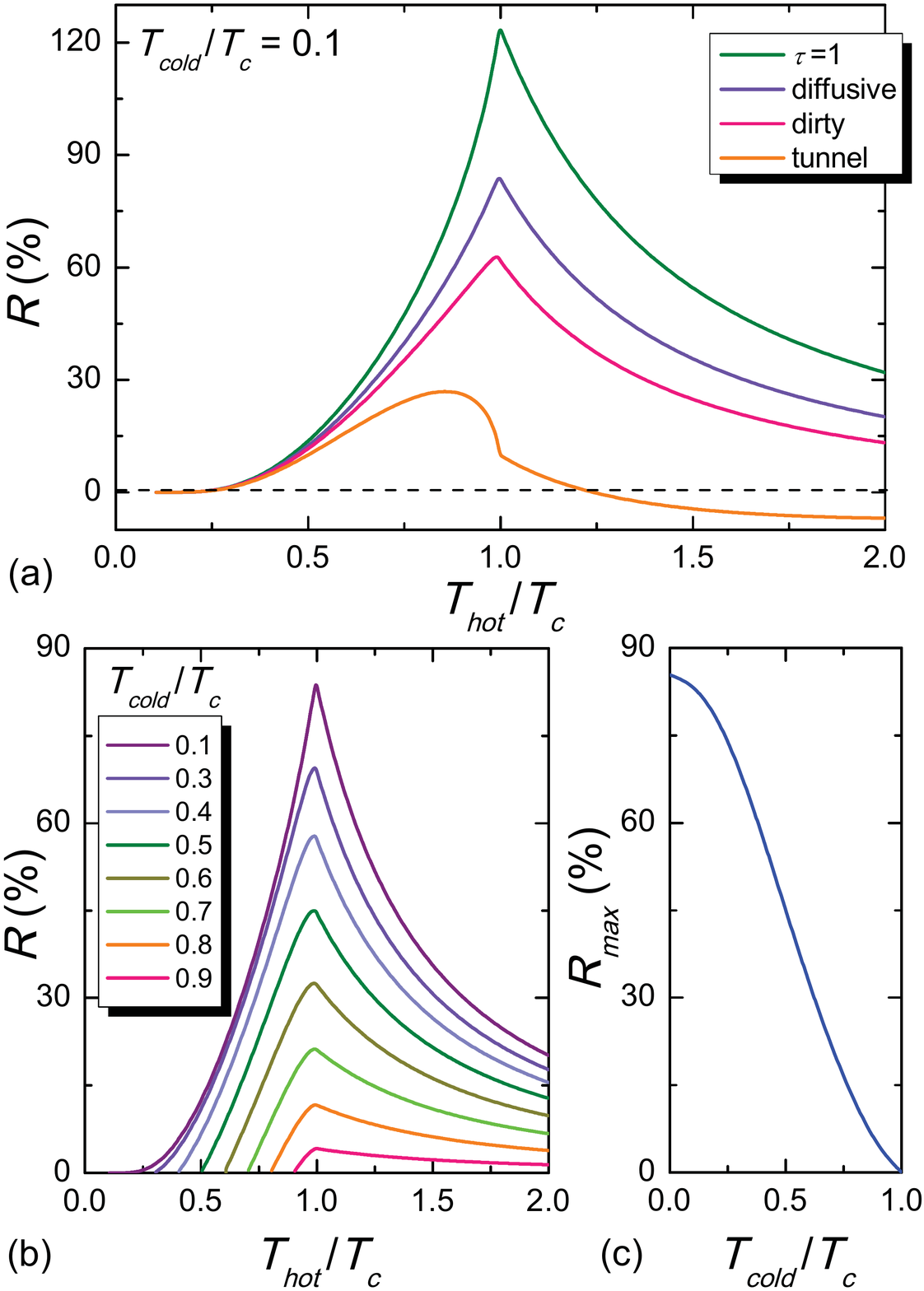} 
\caption{(a) Comparison of thermal rectification coefficient $R$ vs $T_{hot}$ for four different types of NS junctions calculated at $T_{cold}=0.1 T_c$.
(b) $R$ vs $T_{hot}$ for a \emph{diffusive} NS junctions calculated for several values of $T_{cold}$.
(c) Maximum rectification efficiency $R_{max}$ vs $T_{cold}$ for a diffusive NS junction.
}
\label{fig3}
\end{figure}

We are now able to explore the thermal diode properties of the NS contact by calculating the rectification coefficient [see Eq.(\ref{rect})].
To this end it is illustrative to start our discussion considering first the heat rectification characteristics of a  point-contact ballistic NS junction characterized by a unique channel of transmission $\tau$. 
Figure \ref{fig2}(a) shows    the rectification efficiency $R$ vs $T_{hot}$ for $T_{cold}=0.1 T_c$ and for several values of $\tau$. 
Above, $T_c=(1.764k_B)^{-1}\Delta_0$ is the superconducting critical temperature while $\Delta_0$ is the zero-temperature energy gap.
In general, for any  transmission, $R$ is a non-monotonic function of the temperature peaked at a specific $T_{hot}$ which depends on $\tau$, then rapidly decreasing at higher temperature.
In particular, for a perfect transmissive interface  ($\tau=1$) a maximum  thermal rectification coefficient as high as $\sim 123\%$ is obtained at $T_{hot}=T_c$.
This large $R$ value  stems from ideal Andreev reflection \cite{Andreev1964} at the NS interface. 
For $\tau \gtrsim 0.1$ the heat rectification turns out to be always \emph{positive} in the whole range of temperatures. 
By reducing $\tau$ yields a suppression of the maximum of $R$ which is attained for smaller values of $T_{hot}$. 
Notably,  \emph{negative} $R$ values are obtained at large $T_{hot}$ temperatures, i.e., for $T_{hot}>T_c$. 
This sign inversion of the thermal rectification coefficient implies that the heat current flows preferentially  from N to S. 
For low interface transmissivity (i.e., $\tau =10^{-4}$), which describes a NIS tunnel junction, $R$ reaches values as large as $\sim 26\%$ at $T_{hot}\simeq 0.85 T_c$ \cite{MartinezRect2013}. 
We stress that the latter value is around $\sim 20\%$ of the maximum  reached in the  the $\tau=1$ limit. 
It is worthwhile to mention that thermal rectification is a fully non linear effect, and that it is absent in the linear response regime. 
The dependence of the maximum   thermal rectification efficiency ($R_{max}$) as a function of   the transmission coefficient for a point-contact  is shown in Fig. \ref{fig2}(b).  In particular, the  plot shows that for $\tau=0.5$ thermal rectification is reduced by almost a factor of two  with respect to the ideal junction, whereas the lowest saturation limit is already reached for $\tau\lesssim 10^{-3}$.

The effect of the smaller temperature $T_{cold}$ onto $R$ for a perfectly-transmitting NS point-contact is displayed in Fig. \ref{fig2}(c) as a function of  $T_{hot}$. In particular, the increase of  $T_{cold}$ leads to a suppression of $R$.
We emphasize that  the sign of thermal rectification turns out to be  positive in the whole range of temperatures, while $R$ obtains its maximum values always for $T_{hot}=T_c$. 
The evolution of the maximum rectification efficiency $R_{max}$ with $T_{cold}$ is shown in Fig. \ref{fig2}(d). It can be noted how $R$ it is reduced by increasing the temperature. In particular, $R$ reaches $\sim 57\%$ of the maximum at $T_{cold}=0.5 T_c$.

In order to  assess  the  full applicability of heat rectifiers based on NS junctions we consider now less ideal hybrid contacts, i.e., 
NS junctions with  diffusive or dirty interfaces. These are 
 characterized by distributions of transmissivities described by Eqs. (\ref{diffusive}) or (\ref{dirty}), respectively. 
Figure \ref{fig3}(a) shows the comparison of the thermal rectification coefficient $R$ versus $T_{hot}$ calculated at $T_{cold}=0.1T_c$ for four representative different types of NS interfaces: ballistic ($\tau_n=1$), diffusive, dirty and tunnel ($\tau_n\ll1$).   
In particular, for  diffusive and dirty interfaces  $R$ turns out to be always positive, with a shape strongly resembling that of the ballistic case. 
The maximum values for $R$ are  $\sim 84\%$ and $\sim 63\%$ for a diffusive and dirty interface, respectively, and occur at $T_{hot}=T_c$ . 
Such a reduction of the $R$ coefficient stems from a substantial suppression  of the Andreev reflection transmission occurring  in diffusive or dirty contacts in comparison to the fully-transmitting ballistic case \cite{Belzig2000}.
In spite of such a reduction,   both  diffusive and dirty  junctions are  still able to provide a sizeable thermal rectification efficiency which obtains values up to  factor of $3$ larger  than the maximum achievable with a NIS tunnel junction. 

In Fig. \ref{fig3}(b) we show the behavior of $R$ for a diffusive NS junction calculated against $T_{hot}$ for several values of $T_{cold}$ (for dirty interfaces similar results, not shown here, are obtained).
By increasing $T_{cold}$ yields a reduction of the maximum rectification efficiency, being the sign always positive. 
The dependence of $R_{max}$ on $T_{cold}$ is displayed in Fig. \ref{fig3}(c), and shows that at $T_{cold}=0.5 T_c$ the coefficient $R$ can obtain values as large as the $\sim 54 \%$ of the maximum achievable. 
The behavior described above for a diffusive NS contact therefore confirms the picture that  this kind of junctions can provide a substantially large $R$ in a wide range of temperatures. 

From a practical point of view and in light of a realistic implementation, superconducting aluminum (Al) or vanadium (V) combined, for instance,  with copper (Cu) as a normal metal would allow the fabrication of diffusive NS nanojunctions \cite{Ronzani2013,Garcia2009,Courtois2008}. 
On the other side, InAs-based two-dimensional electron gases combined with niobium (Nb) would enable the realization of Schottky barrier-free highly-transmissive semiconductor-superconductor ballistic junctions \cite{Amado2013,Carillo2006,Giazotto2004}.
These predictions for thermal rectifications could be tested experimentally in a prototype hybrid microstructure designed along the lines of that presented in Ref. \cite{MartinezRect2013}, 
in particular by symmetrically tunnel-coupling two additional identical normal metal electrodes to the NS junction. 
Electron heating and thermometry can be performed through NIS tunnel or SNS Josephson junctions \cite{Giazotto2006} coupled to the N leads, therefore allowing to realize selectively the forward and reverse thermal-bias configuration in the structure.
Concerning   potential applications, NS thermal rectifiers could be exploited in the field of electronic cooling,  or for thermal isolation and heat management at the nanoscale. Moreover, other caloritronic devices such as heat interferometers,  sensitive radiation detectors or magnetic sensors might likely benefit from the combination with NS thermal diodes to improve their performance. 

In summary, we have theoretically analyzed thermal rectification in normal metal-superconductor nanojunctions comparing different types of NS contacts. 
We have shown, in particular, that by increasing the interface transmissivity leads to a substantial enhancement of the heat diode effect whereas the sign of rectification can be changed in a suitable range of temperatures for low junction transparency.
For  perfectly-transmissive ballistic contacts, thermal rectification can be as high as $123\%$ thus exceeding by a factor close to $\sim 5$ that achievable in NIS tunnel junctions. For diffusive contacts, the rectification efficiency can obtain values as high as $\sim 84\%$. 
Because of the above results and of the ease intrinsic in their fabrication, NS junctions appear therefore  as a promising building block for the implementation of effective heat diodes to be exploited in low-temperature caloritronic nanocircuitry.

F.G. acknowledges the Italian Ministry of Defense through the PNRM project ``TERASUPER'', and the Marie Curie Initial Training Action (ITN) Q-NET 264034
for partial financial support. 
The work of F.S.B  was supported by the Spanish Ministry of Economy
and Competitiveness under Project FIS2011-28851-C02-02. F.S.B thanks Prof. Martin
Holthaus and his group for their kind hospitality at the Physics Institute of the 
Oldenburg University.

\end{document}